\documentstyle[aps,prl,twocolumn,psfig]{revtex}

\draft
\begin{document}
\newcommand {\be}{\begin{equation}}
\newcommand {\ee}{\end{equation}}
\newcommand {\bea}{\begin{eqnarray}}
\newcommand {\eea}{\end{eqnarray}}
\newcommand {\nn}{\nonumber}

\twocolumn[\hsize\textwidth\columnwidth\hsize\csname@twocolumnfalse%
\endcsname

\title{
Universal Spin-Flip Transition in Itinerant Antiferromagnets}

\author{Georgios Varelogiannis}                            
\address{Department of Physics, National Technical University of Athens,
GR-15780 Athens, Greece}                

\date{\today}
\maketitle

\begin{abstract}

We report a universal spin flip (SF) transition as a function
of temperature in spin-density-wave (SDW) systems. At low temperatures
the antiferromagnetic (AFM) polarization is parallel 
to the applied field and above a critical
temperature the AFM polarization
{\it flips} perpendicular to the field. This transition occurs
in {\it any} SDW system and may be considered as a
qualitative probe of the itinerant character of AFM in a given material.
Our SF transition resolves the longstanding
puzzle of the
SF transition observed in cromium and may be at the origin of
the equally puzzling SDW-I to SDW-II transition in Bechgaard salts for 
which we make experimental predictions.

\end{abstract}
]

The study of itinerant antiferromagnetism (AFM) 
started in the early 
fifties when this state has first been observed in Cromium
\cite{Shull} and since then it has been a field of
continuous interest related to some of the most
fascinating problems in materials        
physics. The first consistent theoretical scheme      
for itinerant AFM has been elaborated by Overhauser
\cite{Overhauser} who introduced the spin density wave
(SDW) picture. The itinerant character of AFM  
and the relevance of the
SDW picture in cromium            
are firmly established experimentally
\cite{Stassis,Fawcett}. 
However,
despite several decades of intense
theoretical work which led to the construction 
of a successful microscopic SDW model for cromium
\cite{Overhauser,Fedders,Shibatani,Rice,Nakanishi,Kotani,Fenton,Machida,Fishman},
there is a surprising aspect of the AFM behavior in this material
which escapes any microscopic understanding so far.
It is the famous {\it spin-flip} (SF) transition as a function of
temperature for which there are only phenomenological
accounts within a Landau framework 
\cite{Fawcett,Shimizu,Walker,Buker}.  
Spin-orbit coupling and
dipole-dipole interactions have been shown to
be unable to produce a spin-flip transition
with temperature \cite{Allen}.
Lacking any understanding of this first order SF transition,
it is viewed up to now as a mysterious 
peculiarity of cromium.                                     

Other very 
extensively studied SDW materials 
are the so called Bechgaard salts which 
attracted much interest not only for their SDW behavior,
but also for its interplay with superconductivity
and related exotic 
phenomena like field-induced SDW transitions     
and quantum-Hall-effect phenomena   
\cite{Jerome}. These salts are quasi-one-dimensional
organic compounds having the form $(TMTSF)_2-X$ where
X denotes a monovalent ion and TMTSF is for
tetramethyltetraselenafulvalene \cite{Bechgaard}.
It has been established recently that {\it inside
the SDW phase} there is a surprising transition
to a {\it new} SDW phase \cite{Takahashi,Valfells}.
This {\it SDW-I to SDW-II} transition manifests 
by a sudden change in the 
T-behavior of the NMR relaxation rate from linear
just below $T_{Neel}$ to an exponential Arhenius behavior
at lower temperatures \cite{Takahashi,Valfells}. 
So far, this phenomenon has been regarded as a peculiar   
transition from an incomplete SDW state below $T_N$
to a complete SDW state at low T, but such a picture is not consistent
with the transport behavior. A spin glass transition has
also been invoked \cite{Lasjaunias}. 

In the present Letter we point out that {\it the spin-flip transition
observed in cromium
is in fact a totally generic phenomenon in
itinerant AFM}. 
We show that the
zero temperature field-induced {\it spin-flop} transition
in itinerant AFM, at finite
temperatures it occurs at a lower critical field and
at a sufficiently high temperature 
it occurs at an {\it arbitrarilly small field}
giving rise to the {\it T-induced spin-flip transition}.
This SF transition is absolutely generic characterizing {\it any 
SDW state} and therefore should manifest in {\it all
itinerant antiferromagnets} when crystal fields are
negligible. Such a generic
behavior of the SDW state has not been noticed so far probably
because in most theoretical works on SDW a one-dimensional
framework was adopted lacking the extra spatial dimensions
involved in the SF transition.
All aspects of the well studied SF transition
in cromium \cite{Fawcett} are in agreement with our SF transition.     
As for the {\it SDW-I to SDW-II} transition in TMTSF's,
we {\it predict} the identification of a similar SF transition 
as the one observed in cromium
when the measurements of Ref. \cite{Mortensen}
will be extended to lower temperatures.
In fact we argue that the available NMR data
\cite{Takahashi,Valfells} are totally compatible with our SF transition.
Our SF transition can be regarded  
as a qualitative probe
of the itinerant SDW character of AFM in a given material.

We consider the most general mean-field Hamiltonian
describing a SDW state in                   
the presence of a uniform
magnetic field:
\bea
H = \sum_{\bf{k},\alpha}\xi_{\bf{k}\alpha}
c^{\dagger}_{\bf{k}\alpha}c_{\bf{k}\alpha} 
+\mu_B\sum_{\bf{k},\alpha,\beta}c^{\dagger}_{\bf{k}\alpha}
\bigl( {\bf{\sigma}}\cdot {\bf H}\bigr) c_{\bf{k}\beta}
\cr
-\sum_{\bf{k},\alpha,\beta}(\bf{\sigma}\cdot \bf{n})_{\alpha\beta}
M_{\bf{k}}\biggl(
c^{\dagger}_{\bf{k}\alpha}c_{\bf{k}+\bf{Q}\beta} + {\it hc}\biggr)
\eea
where $\alpha,\beta$ index the spin, $M_{\bf{k}}$ is the SDW
order parameter, {\bf n} defines the axis of the magnetic 
polarization of the SDW   
and {\bf H} the Pauli contribution of the applied
magnetic field. Orbital effects of the magnetic field are
irrelevant in the SDW state.
The electronic
dispersion $\xi_{\bf k}$ considered in the numerical calculations
reported
here is a tight-binding scheme for a square two dimensional 
lattice with nearest-neighbors hoping 
$\xi_{\bf k}=t(\cos k_xa + \cos k_ya)$. However our results are
independent of the choice of the dispersion 
as we have verified numerically and discuss later.               

To allow for any relative orientation between the
SDW polarization and the direction of the 
field we will use an eight-component spinor
formalism. This eight component space
is overcomplete for the present problem, however, it 
allows to consider elsewhere the same 
phenomena in the presence of additional   
order parameters \cite{PRLCMR}   
avoiding a problem dependent
formalism. Our space is defined by the eight component
spinor
\bea
\Psi^{\dagger}_{\bf{k}}=\bigl(
c^{\dagger}_{\bf{k}\uparrow}c^{\dagger}_{\bf{k}\downarrow}
c_{-\bf{k}\uparrow}c_{-\bf{k}\downarrow}
c^{\dagger}_{\bf{k}+\bf{Q}\uparrow}c^{\dagger}_{\bf{k}+\bf{Q}\downarrow}
c_{-\bf{k}-\bf{Q}\uparrow}c_{-\bf{k}-\bf{Q}\downarrow}\bigr)
\eea
The following tensor products provide a convenient basis
for the projection of the Hamiltonian
in this spinor space
\bea
\widehat{\tau}_i=\widehat{\sigma}_i\otimes\bigl(\widehat{I}\otimes 
\widehat{I})\nn \\       
\widehat{\rho}_i=\widehat{I}\otimes\bigl(\widehat{\sigma}_i\otimes
\widehat{I})\\             
\widehat{\sigma}_i=\widehat{I}\otimes\bigl(\widehat{I}\otimes
\widehat{\sigma}_i)\nn
\eea
where $\widehat{\sigma}_i$ are Pauli matrices in usual notations
and $I$ the $2\times2$ identity matrix.
This type of multicomponent formalism has been used 
for the study of  
magnetic superconductors
\cite{Levin} and recently for the study of excitonic         
ferromagnetism and colossal magnetoresistance \cite{PRLCMR}.

When                            
${\bf H}\parallel {\bf n}$ our hamiltonian (1) can be 
written in the eight component spinor space  
as follows:
\bea 
\widehat{H}_{\parallel} = \sum_{\bf{k}}\Psi^{\dagger}_{\bf{k}}\biggl(
\xi_{\bf{k}}\widehat{\tau}_3\widehat{\rho}_3
-M_{\bf{k}\parallel}\widehat{\tau}_1
\widehat{\rho}_3\widehat{\sigma}_3
+\mu_B H \widehat{\rho}_3 \widehat{\sigma}_3\biggr) \Psi_{\bf{k}}
\eea
The Green's function corresponding to this Hamiltonian
is now an $8\times 8$ matrix which
can be shown to take the following form in our                     
representation:
\bea
\widehat{G}_{\parallel} ( {\bf k} , i \omega_n ) =
-\lbrack
i\omega_n+
\xi_{\bf{k}}\widehat{\tau}_3\widehat{\rho}_3
-M_{\bf{k} \parallel}\widehat{\tau}_1
\widehat{\rho}_3\widehat{\sigma}_3
+\mu_B H \widehat{\rho}_3 \widehat{\sigma}_3\rbrack\nn
\\  
\lbrack \omega^2_n+\xi^2_{\bf{k}}+M^2_{\bf{k}\parallel}
+\mu^2_B H^2 - 2\xi_{\bf{k}}\mu_B H \widehat{\tau}_3
\widehat{\sigma}_3
+2M_{\bf{k}\parallel}\mu_B H \widehat{\tau}_1
\rbrack\nn
\\  
\lbrack\omega^2_n+E^2_{+\parallel}({\bf k})\rbrack^{-1}
\lbrack\omega^2_n+E^2_{-\parallel}({\bf k})\rbrack^{-1}   
\qquad\qquad
\eea
where 
\bea
E_{\pm\parallel}({\bf k})=
\sqrt{\xi^2_{\bf{k}}+M^2_{\bf{k}\parallel}}\pm \mu_B H
\eea
The SDW gap equation results from the 
self-consistency relation 
$M_{{\bf k}\parallel}={1\over 8}T\sum_{\bf k'}\sum_n
V_{{\bf k k'}}\times
Tr\{\widehat{\tau}_1
\widehat{\rho}_3\widehat{\sigma}_3
\widehat{G}_{{\bf k'}n\parallel}\}$
and after analytic summation over the Matsubara frequencies it
can be shown to take the following form
\bea
M_{{\bf k}\parallel}=\sum_{{\bf k'}}
V_{{\bf k k'}}
{M_{{\bf k'}\parallel}\over 4\sqrt{\xi^2_{\bf{k'}}+
M^2_{{\bf k'}\parallel}}}
\biggl[\tanh \biggl( {E_{+\parallel} ({\bf k'})\over 2T}\biggr)
\cr +
\tanh \biggl( {E_{-\parallel} ({\bf k'})\over 2T}\biggr)\biggr]
\eea
which is identical with the gap
equation of
a singlet BCS superconductor in a Zeeman field.
The field $\mu_B H$ appears only in the hyperbolic tangent
functions and 
in the 
zero temperature regime we have
$|\tanh(E_{\pm\parallel}({\bf k})/2T)\approx 1|$.
{\it 
Therefore
a magnetic
field smaller than the critical field
and parallel to the polarization of the SDW has practically no
influence on the SDW in the zero temperature
regime}. On the other hand, when the field is sufficiently large        
$\mu_B H > M_{{\bf k}\parallel}$ then in the $T\rightarrow 0$ regime
$\tanh(E_{-\parallel}({\bf k})/2T)=
-\tanh(E_{+\parallel}({\bf k})/2T)=-1$
and the SDW is eliminated.
Therefore, in the $T\rightarrow 0$ regime there is
a critical magnetic field
parallel to the polarization of the SDW 
($\mu_B H_c\approx M_{{\bf k}\parallel}$ if the gap
is isotropic) that can melt the SDW.
This is the analog of the well known  
Clogston-Chandrasekhar critical field
\cite{Clogston} in superconductivity which has indeed
been observed in superconducting films when the 
field is applied parallel to the film planes \cite{FuldeR}.

\begin{figure}[h]
\centerline{\psfig{figure=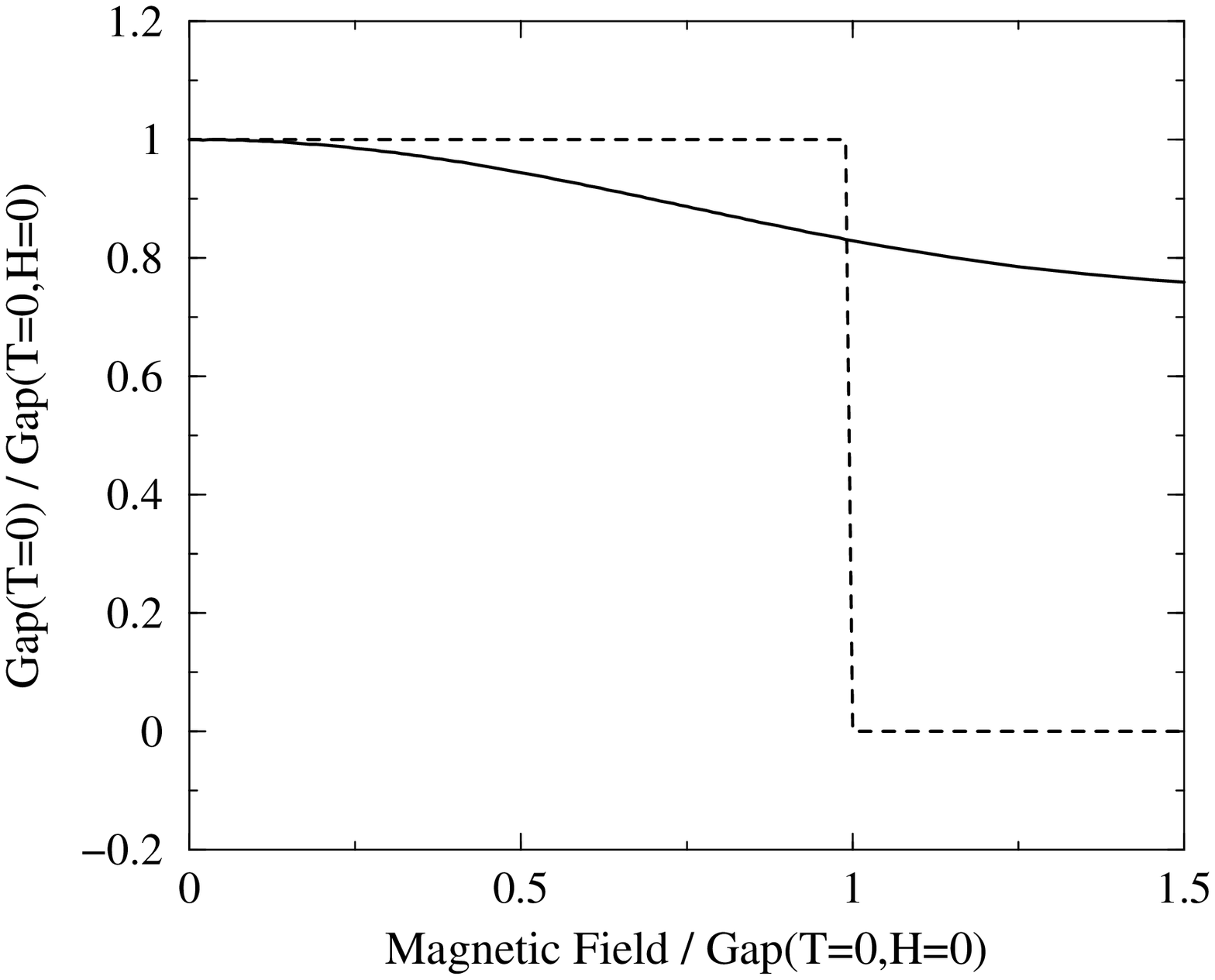,width=8cm,height=5.5cm,angle=0}}
\centerline{\psfig{figure=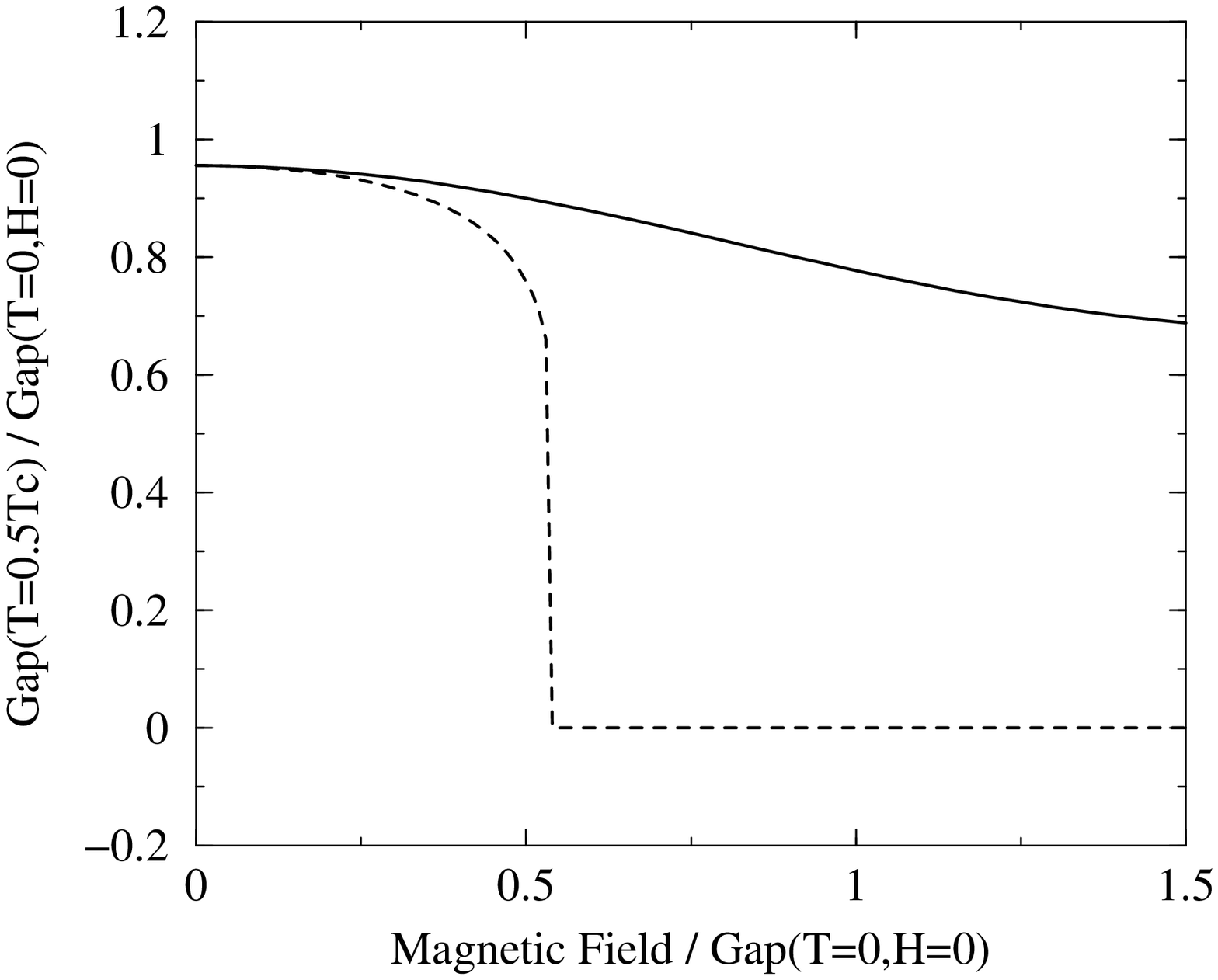,width=8cm,height=5.5cm,angle=0}}
\vspace{0.3cm}
\caption{
a) Evolution of the SDW gap as a function
of the  magnetic field in the zero temperature regime
when the SDW polarization is parallel (dashed line) or
perpendicular (full line) to the field direction.
At low fields the parallel polarization has a lower
free energy (higher SDW gap). When $\mu_BH$ exceeds the gap,
only the perpendicular polarization has a finite gap
leading to the {\it spin-flop} transition from
parallel (at low fields) to perpendicular SDW polarization
as a function of the field. b) Same as in a) but at
a finite temperature $T=0.5T_N$. The {\it spin-flop
transition is eliminated}!
}
\end{figure}

Numerical solutions of the gap equation 
confirm this behavior (see Fig. 1a). Indeed, in the zero temperature
regime $M_{{\bf k}\parallel}$ as a function of the field
has a step-like behavior and for $\mu_B H>M_{{\bf k}\parallel}$
(in the example shown in Fig. 1a the gap is isotropic), 
$M_{{\bf k}\parallel}$
is identically zero.
The melting of the SDW when $\mu_BH_c =M_{{\bf k}\parallel}$
manifests already in the structure of the poles
of the Green's function reported in (6).
One of the two quasiparticles poles 
$E_{-\parallel}({\bf k})$
moves to zero when $\mu_BH_c =M_{{\bf k}\parallel}$ and there is no gap
on the Fermi surface.
Because in $E_{-\parallel}({\bf k})$ the field 
$\mu_B H$ and the SDW gap $M_{\bf{k}}$
contribute into terms which have {\it opposite sign} we can say that
the ${\bf H}\parallel {\bf n}$ magnetic field is 
{\it in direct competition} with the SDW.
The situation will be shown below to be different if the 
polarization of the SDW is perpendicular to the field.

We now consider the case is which 
${\bf H}\perp {\bf n}$.
In the same eight-component formalism our                  
Hamiltonian (1) can be written as follows:                              
\bea
\widehat{H}_{\perp} = \sum_{\bf{k}}\Psi^{\dagger}_{\bf{k}}\biggl(
\xi_{\bf{k}}\widehat{\tau}_3\widehat{\rho}_3
-M_{\bf{k}\perp}\widehat{\tau}_1
\widehat{\rho}_3\widehat{\sigma}_3
+\mu_B H \widehat{\rho}_3 \widehat{\sigma}_1\biggr) \Psi_{\bf{k}}
\eea
The corresponding matrix Green's fucntion
diagonalized in our representation takes the following form  
\bea
\widehat{G}_{\perp} ( {\bf k} , i \omega_n ) =
- \lbrack
i\omega_n+
\xi_{\bf{k}}\widehat{\tau}_3\widehat{\rho}_3
-M_{\bf{k}\perp}\widehat{\tau}_1
\widehat{\rho}_3\widehat{\sigma}_3
+\mu_B H \widehat{\rho}_3 \widehat{\sigma}_1\rbrack\nn
\\ 
\lbrack \omega^2_n+\xi^2_{\bf{k}}+M^2_{\bf{k}\perp}
+\mu^2_B H^2 - 2\xi_{\bf{k}}\mu_B H \widehat{\tau}_3
\widehat{\sigma}_1
\rbrack\nn
\\ 
\lbrack\omega^2_n+E^2_{+\perp}({\bf k})\rbrack^{-1}
\lbrack\omega^2_n+E^2_{-\perp}({\bf k})\rbrack^{-1}
\eea
and the quasiparticles poles are now defined by
\bea
E_{\pm\perp}({\bf k})=
\sqrt{(\xi_{\bf{k}}\pm \mu_B H)^2+M^2_{\bf{k}\perp}}
\eea

From the structure of the poles 
it is already obvious that 
the perpendicular field {\it is not in direct competition 
with the SDW}. None of the quasiparticle poles given in (10)
can be set to zero no matter how large the magnetic field is.
This indicates that the magnetic field cannot melt the SDW.
Let us check this by calculating the gap equation which
can now be shown to take the following form:
\bea
M_{{\bf k}\perp}=\sum_{{\bf k'}}
V_{{\bf k k'}}M_{{\bf k'}\perp}\biggl[
{1\over 4E_{+\perp} ({\bf k'})}
\tanh \biggl( {E_{+\perp} ({\bf k'})\over 2T}\biggr)
\cr+
{1\over 4E_{-\perp} ({\bf k'})}
\tanh \biggl( {E_{-\perp} ({\bf k'})\over 2T}\biggr)
\biggr]
\eea
Only in the limit 
$\mu_B H\rightarrow\infty$ the gap equation provides an identically
zero solution. Moreover, any finite perpendicular field reduces 
gradually the SDW gap (because it
appears in the denominator) {\it even in the $T\rightarrow 0$
regime no matter how small it is} in sharp contrast 
with the parallel field behavior where in the
$T\rightarrow 0$
regime fields smaller than the gap have practically no influence.

The above behavior in the $T\rightarrow 0$ regime is verified
by numerical solutions as shown in Fig. 1a.
Therefore, if the polarization ${\bf n}$ of the SDW is
free as in any perfectly itinerant SDW system,
we naturally expect the following behavior of ${\bf n}$
in the presence of a field in the $T\rightarrow 0$ regime.
{\it For weak fields the polarization of the SDW will chose
the direction parallel to the field since in 
that way it is insensitive on it}. As the field grows,
and as long as it remains smaller than the gap, ${\bf n}$ remains
locked parallel to the direction of the field.
When the field equals the gap, the SDW will {\it suddenly
flip its polarization from ${\bf n \parallel H}$ to 
${\bf n \perp H}$}. This {\it first order} transition
illustrated in Fig. 1a is the itinerant counterpart
of the well studied {\it spin-flop} transition in the localized magnetism
picture. However, the situation is qualitatively different
here. In fact, in the localized magnetic moments picture,
at any finite field
the moments have tendency to be perpendicular
to the field while here this tendency exists only above the 
critical field.

\begin{figure}[h]
\centerline{\psfig{figure=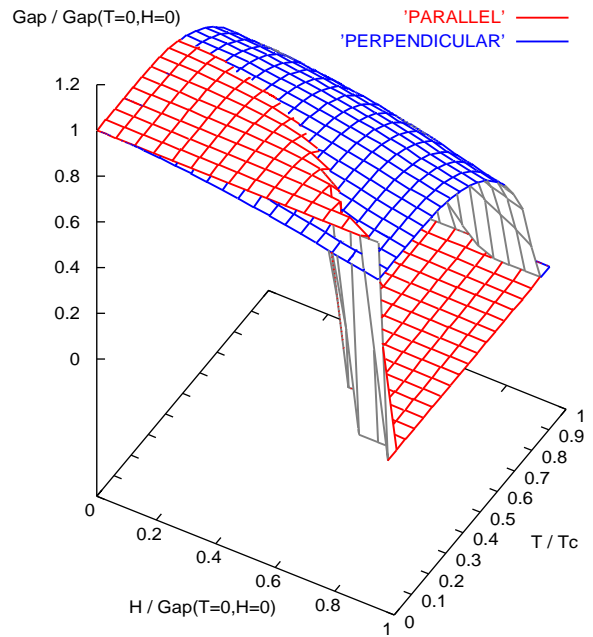,width=8cm,height=10.0cm,angle=0}}
\vspace{0.3cm}
\caption{
(color): Evolution of the SDW gap as a function
of the  magnetic field and temperature for parallel
(red) and perpendicular (blue) to the field polarization of the SDW.
At low fields and temperatures the parallel polarization prevails.
}
\end{figure}

More suprising, and without counterpart in the localized
limit, is the behavior of our {\it spin-flop} transition 
at finite temperatures. Finite temperature solutions of the 
gap equations indicate
that although at low temperatures and low
fields the ${\bf n\parallel H}$ state prevails (see Fig. 1a), at a higher
temperature $T\approx T_N/2$ the {\it ${\bf n\perp H}$ state
prevails whatever the field} (Fig. 1b). 
Above a given temperature, the magnetic field induced spin-flop
transition is in fact eliminated.
An example of the evolution 
of both $M_{\parallel}$ and $M_{\perp}$ as 
a function of the field and the temperature is reported
in Fig. 2. 
The zero temperature {\it spin flop} transition
from ${\bf n \parallel H}$ to ${\bf n \perp H}$,
by rising the temperature it appears
at lower critical fields and at $T\approx 0.435 T_N$
the {\it critical field of this transition is zero}.
The physical origin of this thermally induced spin-flip transition
is probably related with the phase space for 
spin fluctuations. In fact, with the SDW polarization
perpendicular to the field, the available phase space
for thermal excitation of the spins is larger than in the case of
a polarization parallel to that of the field. At a sufficiently high temperature
this phase-space gain apparently dominates inducing the
spin-flip transition.

Our SF transition displays as a function
of the field and the temperature {\it all the characteristics
of the SF transition in Cromium} \cite{Fawcett} 
which we believe is its most obvious 
physical realization.
This is further supported by the fact that the order
of magnitude of the ratio $T_{SF}/T_N$ in cromium is just in the 
range in which we predict this transition should happen.
Moreover, when particle-hole asymmetry is introduced including
for example a next nearest neighbors hopping term in our
dispersion and reducing thus the nesting, our $T_{SF}/T_N$
is reduced and {\it this precisely what is observed by   
alloying cromium} \cite{Fawcett}. 
The most likely range of this transition is
$0.20T_N\leq T_{SF}\leq 0.45T_N$, the highest value being
indicative of a particle-hole symmetric system.
In bare cromium 
$T_{SF}\approx 0.395 T_N$ which is just in the range where
we expect our SF transition.

As for the {\it SDW-I to SDW-II} transition in Bechgaard salts,
here as well it happens precisely in the temperature range in 
which we predict our SF transition ($\approx T_{Neel}/3$).
Moreover, the Arhenius low-T behavior of the NMR relaxation rate
\cite{Takahashi,Valfells} is consistent with {\bf n} locked parallel
to the field while the linear Korringa behavior at higher T
and up to $T_{Neel}$ is consistent with both {\bf n} 
perpendicular to the field and the observed insulating
transport behavior. We predict that extending the measurements 
of
\cite{Mortensen} to temperatures below 4K could
definitely establish the
SF character of the {\it SDW-I to SDW-II} transition which is observed
at about 3.5 K in Bechgaard salts.

In conclusion, we have shown that in {\it all} SDW systems 
occurs a SF transition by varying temperature.
In the low-T phase    
the SDW polarization is parallel to the field while above
the SF transition it is perpendicular. This SF transition
has been identified in cromium and is likely to be the origin of 
the {\it SDW-I to SDW-II} transition in Bechgaard salts.
It represents a qualitative tool for identifying the 
itinerant character of AFM in a given material.

Valuable
discussions with Michel H\'{e}ritier, Denis Jer\^{o}me, Peter Littlewood,
Peter Oppeneer, Nikos Papanikolaou and 
Peter Thalmeier are gratefully acknowledged.


\end{document}